http://orcid.org/0000-0003-4994-0573

# Qbits, la función de onda del universo, diagramas de Penrose: Una nueva perspectiva para una física simbólica (I)

Martín Pomares Calero
Docente del Departamento de Física
Facultad de Ciencias e Ingeniería

**Resumen**

La presente investigación trata de un análisis acerca de la función de onda del universo la cual según la mecánica cuántica debe contener toda la información de los estados cosmológicos y cuánticos del universo. Así como también debe de incluir la historia de cómo el universo evolucionó en diversos estados considerando el big bang como el estado base de energía de creación. Bajo la consideración entrópica de Hawking para un black hole, la función de onda puede contener tanta información que si colapsa dicha información puede ser transferida en forma de radiación. En este trabajo se trata de analizar a la función de onda y sus estados cuánticos desde el punto de vista de la computación cuántica de tal manera que se plantea el hecho de que la función de onda del universo puede describir los estados cosmológicos en forma de Qubits, de tal manera que la radiación contendría suficiente Qubits como sean necesario para analizar sucesos o anomalías espaciotemporales.

**Introducción**

En mecánica cuántica la función de onda se le considera que encierra toda la información del sistema cuántico que se estudia, en cambio en cosmología cuántica entenderemos a la función de onda representando un paquete de vibraciones o como una corriente cuántica que en nuestro caso todo lo imaginable está hecho de ondas. Estas funciones de onda son holográficas y tienen una geometría precisa la cual es así mismo información. Cada función de onda es un paquete de información [1], es decir contiene toda la información del universo. Dicha información puede ser representada por qbits. El concepto de qbits proviene de quantum-bit de información o bien bit cuántico de información que puede ser almacenada en la materia utilizando los diversos niveles de energía que hay existentes en un átomo. Sin embargo, esto es el tema fundamental de investigación dentro de lo que hoy se conoce como computación cuántica. Todo esto implica que el universo actúa como una computadora cuántica, es decir una computadora holográfica de información ya que cada estado del universo puede ser representado por información. Y en nuestro caso información cuántica. Todo esto es parte también de lo que ahora actualmente se conoce como cosmología cuántica en la cual en algunas investigaciones al universo se le considera como un todo con consciencia, una consciencia que está impregnada en todo el universo que en parte es lo que podemos definir como la inteligencia superior o Dios [2, 3].

**Más a cerca de la función de onda del universo**

Por otra parte, dentro de la cosmología cuántica ha surgido la nueva idea de Hawking en tratar el universo entero como si fuera una partícula cuántica [4]. Para este caso la función de onda del universo sin frontera [5, 6] puede ser descrita como:

$$\Psi = \int \delta g \delta \phi \, e^{-I[g,\phi]}$$

donde $g$ es la métrica, y $\phi$ es un campo escalar. Sin embargo, una condición de frontera que es frecuentemente escogida es la que conduce a la bien conocida la función de onda de Hartle-Hawking (HH) [7]. Si se aplican técnicas estándar de cuantización canónica para relatividad general podemos derivar una ecuación no temporal para la función de onda del universo [8]. Además de considerar las condiciones de Einstein-Bell en las cuales, el universo se debe considerar local [8], lo cual implica que las condiciones de frontera para la función de onda del universo son locales.

La magnitud $\rho(a) = |\psi(a)|^2$ para el universo es inversamente proporcional al parámetro de Hubble $H(a)$ del universo, y representa la densidad de probabilidad del universo estando en un estado $a$ (el factor de escala) durante su evolución [9]. $\rho(a)$ representa la velocidad de evolución del universo y es proporcional al tiempo requerido para que el universo evolucione del estado $a_1$ al estado $a_2$ [9]. La interpretación dinámica de la función de onda del universo puede dar soluciones de inflación de un pequeño universo y la correcta evolución de las leyes del universo en el límite clásico tal como es requerido por el principio de correspondencia [9].

De acuerdo a la computación cuántica los qbits pueden ser representados mediante el producto de estados (producto tensorial) descritos en función de ceros y unos tal como [10]:

$$|0\rangle \otimes |0\rangle, \; |0\rangle \otimes |1\rangle, \; |1\rangle \otimes |0\rangle, \; |1\rangle \otimes |1\rangle,$$

O en su forma reducida

$$|0\rangle|0\rangle, \; |0\rangle|1\rangle, \; |1\rangle|0\rangle, \; |1\rangle|1\rangle,$$

Los cuales pueden reescribirse en una forma más compacta como [10]

$$|00\rangle, \; |01\rangle, |10\rangle, |11\rangle$$

En forma matricial los kets ceros y uno pueden ser representados como

$$|0\rangle \leftrightarrow \begin{pmatrix}1\\0\end{pmatrix}, \quad |1\rangle \leftrightarrow \begin{pmatrix}0\\1\end{pmatrix}$$

En vista que los qbits se representan y operan mediante producto tensoriales es posible desarrollar operaciones más complejas de forma tensorial también, lo que implica que operaciones complejas necesitan una forma de representación más simplificada para la cuantificación de las magnitudes que se intentan representar mediante los estados. En ese sentido podemos utilizar formas simbólicas de cálculo muy similares a los conocidos diagramas de Feynman, pero en nuestro caso se denominan *diagramas tensoriales de Penrose* representados en la Fig. 1. Estos diagramas representan operaciones tensoriales que pueden cuantificar de manera más simplificada las operaciones tensoriales entre magnitudes físicas [11].

Los diagramas tensoriales de Penrose son un indicio para establecer las bases de una nueva álgebra de cálculo para la física la cual según el autor puede ser denominada *física simbólica*. Esta vendría a ser una nueva área de investigación pues sería una herramienta perfecta para investigar la conexión entre el macro y micro mundo, es decir entre lo clásico y lo cuántico.

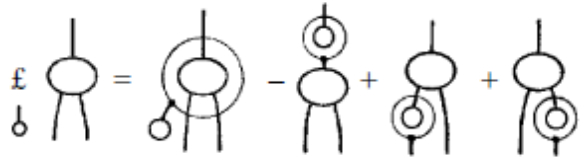

Fig. 1. Diagramas Tensoriales de Penrose. Figura tomada del libro "The Road to Reality: A complete guide to the laws of the universo" escrito por Roger Penrose [11].

Según Hawking, la función de onda del universo puede ser usada también para investigar las condiciones de un universo localmente pequeño como es el caso de un *black hole* el cual irradia información cuántica (Radiación de Hawking) con cierta cantidad de entropía $S$ que depende del área del horizonte de sucesos $A$ descrita como [12]:

$$S = \frac{k_B c^3}{4G\hbar} A$$

Cerca del horizonte de sucesos, las partículas virtuales dentro del black hole absorben energía del campo gravitatorio y se convierten en partículas reales de las cuales algunas escapan del agujero negro (Fig. 2) [11]. Las partículas que escapan pueden ser cuantificadas en función de su estado cuántico en forma de qbits que luego puede ser analizada.

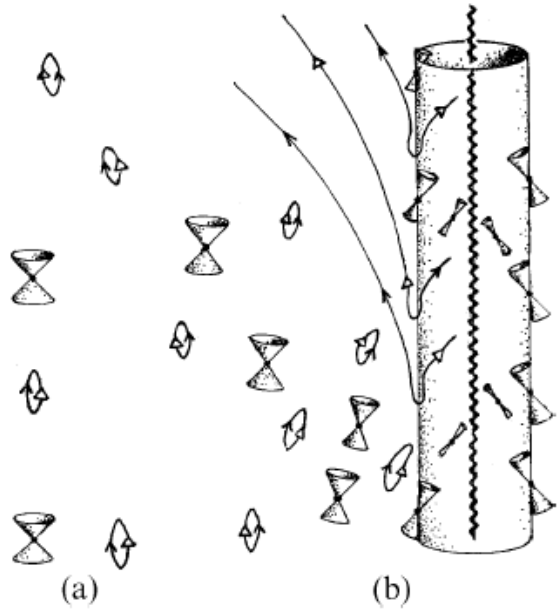

Fig. 2 Representación intuitiva de la radiación de Hawking. (a) Lejano del black hole, el par de partículas-antipartículas virtuales son continuamente producidas, y en corto tiempo son aniquiladas. (b) Muy cercano del horizonte del black hole una de las partículas del par partículas-antipartículas virtuales escapa del black hole convirtiéndose en una partícula real. La ley de conservación de la energía demanda que las partículas que caen en el black hole tengan energía negativa. Figura tomada del libro "The Road to Reality: A complete guide to the laws of the universo" escrito por Roger Penrose [11].

**A manera de Conclusión**

Un Nuevo reto para la física sería el desarrollar la física simbólica a la par de la computación cuántica para desarrollar computadoras cuánticas que operen de manera tensorial, o bien desarrollar software que realice las operaciones tensoriales de la física simbólica. Este sería una nueva área de investigación totalmente virgen a explorar para futuras generaciones de físicos. Además, dichas computadoras cuánticas desarrolladas para este fin podrían ser herramientas útiles para tratar de unificar el macro y el micro mundo lo cual es parte del santo grial de la física en nuestros días.

**Referencias Bibliográficas**

[1]. Huntley, Noel. The Wave Function is Real: The Holographic Quantum Model. Source: Beyond Duality. http://www.users.globalnet.co.uk/~noelh/Wavefunction.htm

[2]. Walia, Arjun (2014) "Consciousness Creates Reality" – Physicists Admit The Universe Is Immaterial, Mental & Spiritual. Collective Evolution. http://www.collective-evolution.com/2014/11/11/consciousness-creates-reality-physicists-admit-the-universe-is-immaterial-mental-spiritual/

[3]. Stewart-Williams, Steve (2010) Is the Universe Conscious? Humans evolved and the universe woke up. The Nature-Nurture-Nietzsche Blog.

[4]. Kaku, M. () Wave Function of the Universe(s) From *Hyperspace*.

[5]. Hartley, James B. Theories of Everything and Hawking's Wave Function of the Universe. Department of Physics, University of California. Santa Barbara. USA.

[6]. Ivancevic, V. G.,Ivancevic, T. T (2007) Quantum Leap: From Dirac and Feynman, Across the Universe, to Human Body and Mind. World Scientific.

[7]. Grishchuk L. P.; Sidorov, Yu. V. (1988) Boundary conditions for the wave function of the universe.

[8]. Evans, Peter W. et all. (2016) Ψ-Epistemic Quantum Cosmology?

[9]. He, Dongshan et all. (2015) Dynamical interpretation of the wavefunction of the universe.

[10]. Mermin, N. David (2002) From Cbits to Qbits: Teaching Computer Scientists Quantum Mechanics.

[11]. Penrose, Roger (2004) THE ROAD TO REALITY: A Complete Guide to the Laws of the Universe. JONATHAN CAPE, LONDON.

[12]. Alonso Villela, A. V. et all. (2014) Radiación de Hawking. Instituto Tecnológico y de Estudios Superiores de Monterrey. México.